February 2023

# Dataset Management Platform for Machine Learning

Ze Mao

Yang Xu

Erick Suarez

# Dataset Management Platform for Machine Learning

ABSTRACT

The quality of the data in a dataset can have a substantial impact on the performance of a machine learning model that is trained and/or evaluated using the dataset. Effective dataset management, including tasks such as data cleanup, versioning, access control, dataset transformation, automation, integrity and security, etc., can help improve the efficiency and speed of the machine learning process. Currently, engineers spend a substantial amount of manual effort and time to manage dataset versions or to prepare datasets for machine learning tasks. This disclosure describes a platform to manage and use datasets effectively. The techniques integrate dataset management and dataset transformation mechanisms. A storage engine is described that acts as a source of truth for all data and handles versioning, access control etc. The dataset transformation mechanism is a key part to generate a dataset (snapshot) to serve different purposes. The described techniques can support different workflows, pipelines, or data orchestration needs, e.g., for training and/or evaluation of machine learning models.

KEYWORDS

- Dataset management
- Dataset transformation
- Training data
- Data versioning
- Data pipeline
- Data orchestration
- Machine learning

BACKGROUND

In machine learning, a dataset is a collection of data that is used to train and evaluate a model. The quality of the data in a dataset can have a substantial impact on the performance of a machine learning model. Effective dataset management, including tasks such as data cleanup, versioning, access control, dataset transformation, automation, integrity and security, etc., is critical for machine learning researchers and engineers. Effective dataset management can also help improve the efficiency and speed of the machine learning process. However, there are no good tools or processes to manage a dataset, e.g., dataset versions, access control for the dataset, etc., or to transform a dataset. Engineers spend a substantial amount of manual effort and time to manage dataset versions or to prepare datasets for machine learning tasks.

While there are a few related products, they all fall short. DVC [1] is a data version control system, but it focuses on model development and deployment, rather than dataset management or transformation. Deeplake [2, 3] provides the ability to manage datasets, but does not have support for data transformation at a large scale. TFDS [4] does not have versioning and access control. git [5] is a well-known version control solution that is suitable for code or text, but not for large objects. Thus, git is not suitable for machine learning datasets, which are usually very large and often include non-text data.

DESCRIPTION

This disclosure describes a platform to manage and use datasets effectively. The techniques integrate dataset management (version, access control, etc.) and dataset transformation mechanisms. A storage engine is described that acts as a source of truth for all data and handles versioning, access control etc. The dataset transformation mechanism is a key part to generate a dataset (snapshot) to serve different purposes. The described techniques can support different workflows, pipelines, or data orchestration needs.

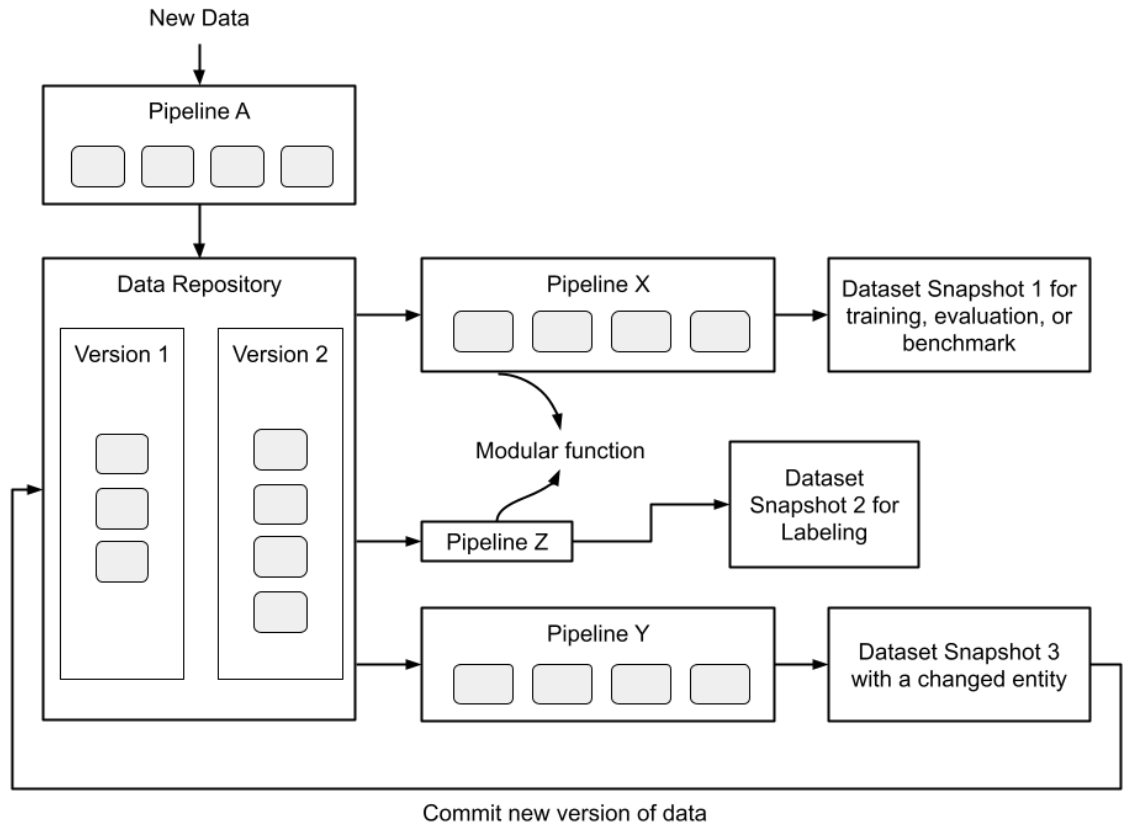

**Fig. 1: Data repository and data pipelines**

As illustrated in Fig. 1, pipelines A, X, Y, and Z are examples of workflows for dataset transformation. New data is ingested via pipeline A and added to the data repository. Each of the pipelines X, Y, and Z generates datasets (snapshots) for different purposes. Data engineers or other users can create different pipelines per their specific needs. For example, snapshot 1 generated by pipeline X can be used to train, evaluate, or benchmark a model, while snapshot 2 generated by pipeline Z can be used for labeling. Pipeline Y creates snapshot 3 which includes at least one changed entity. As shown in Fig. 1, the new version of data in snapshot 3 is committed to the data repository for future use. The pipelines can be triggered manually or automatically based on specific events. The pipelines are maintainable and usable in production environments. The pipelines include the ability of data lineage tracking.

Individual modules in a pipeline (illustrated in Fig. 1as small gray blocks) are shareable, reusable, and chainable. A pipeline operates similar to the extract-transform-load (ETL) pipelines common in big data applications but is more specific to machine learning use cases. A pipeline is lightweight to implement (e.g., is implemented via a few lines of Python code), enables quick iteration, and is easy to run.

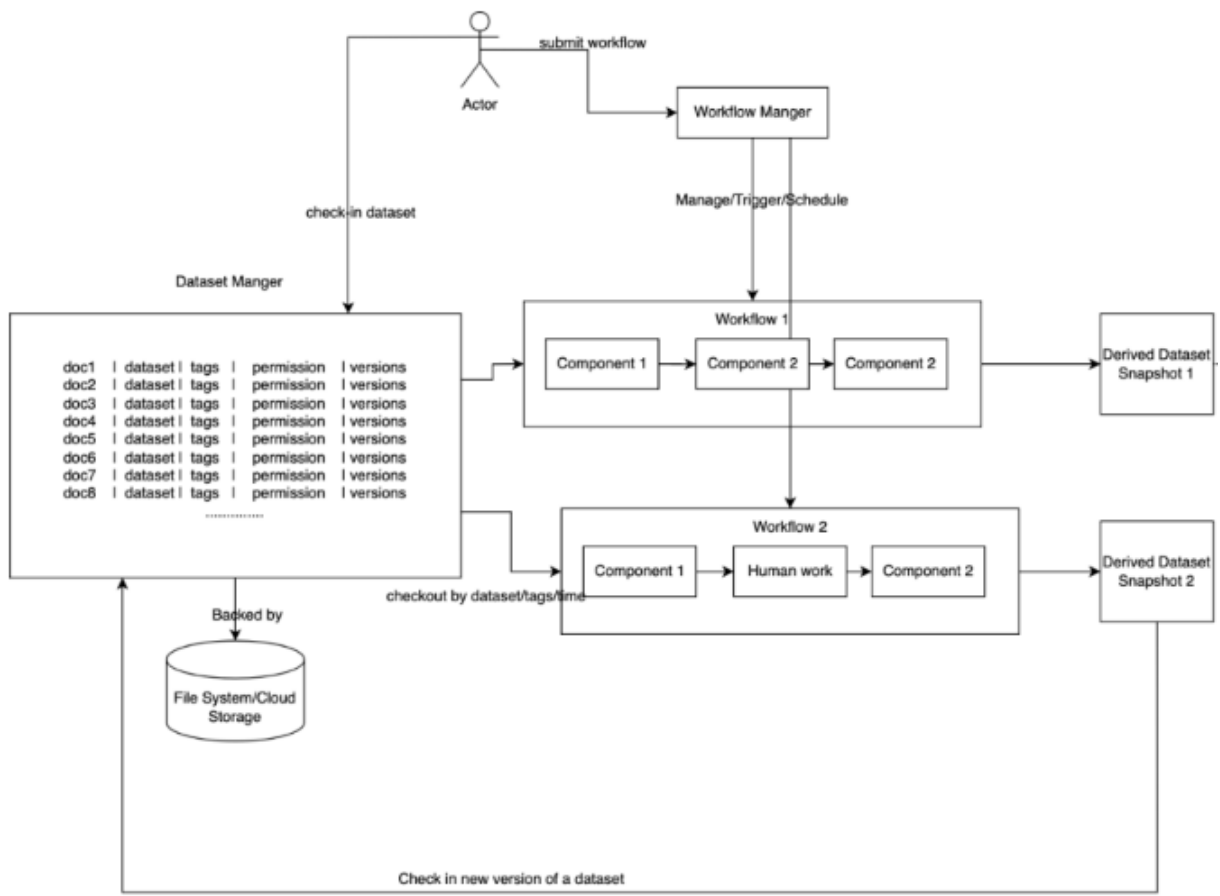

**Fig. 2: Architecture of data management platform**

Fig. 2 illustrates an example architecture of a data management platform, per techniques of this disclosure. An actor (e.g., a user, or an automated program that acts as a trigger) can check-in a dataset and submit a workflow to the platform. The platform includes two core modules: dataset manager and workflow manager.

- **Dataset manager:** The dataset manager is used to store datasets, manage versions, for access control and to checkout datasets. It provides dataset check-in/checkout features. Users can use a command-line interface (CLI) or other user interface (e.g., graphical user interface) to check-in data. Data or datasets can be tagged with one or more tags. The dataset manager enforces access control and permissions at the time of data check-in/checkout. It also provides query capabilities, e.g., querying for datasets by tags, dataset name, or other attributes. Users or workflows can checkout data by specifying query conditions. The type of data stored is unrestricted. Users can check in data in any format. The underlying storage for the data can be any suitable mechanism such as a file system or cloud storage.

- **Workflow manager:** A user that uses the data management platform can register their workflow to the workflow manager. The workflow manager allocates resources, schedules runs, and reports results. In this context, a workflow is a user defined chain of components (pipeline) to transform a dataset. The input to a workflow is a set of checked out data that satisfies particular conditions. The output of a workflow is a set of data that is suitable for in model training or evaluation, or for check-in to the dataset manager as a new dataset. Dataset snapshots derived upon execution of a workflow can be used for training and/or evaluation of machine learning models.

    Each component in a workflow has a functionality. There are two types of components: program based data processing unit and human work based data processing unit. The workflow manager allocates computing resources to the computing

components of a workflow to support large scale data processing. The lineage of data is also tracked.

Key features on the platform include:

- Dataset read/write/delete
- Dataset versioning
    - Version control and version difference
- Dataset access control and security
    - Access control/Permissions
- Dataset transformation
    - Transform the original data to get a derived version of the dataset.
- Workflow automation
    - Orchestrate multiple transformations and human operations into one workflow.
    - Trigger a workflow by event (new dataset version or another event)
    - Schedule workflow by time
- Data lineage
    - Tracking the data lineage by version, derivation, and workflow

The described data management platform can be used in various contexts, such as:

- Performing clean-up operations on newly received data
- Performing data transformations to meet model training/evaluation needs
- Providing automation and connection for programming work and human work based on the data
- Supporting large-scale data processing
- Data revocation

CONCLUSION

This disclosure describes a platform to manage and use datasets effectively. The techniques integrate dataset management (version, access control, etc.) and dataset transformation mechanisms. A storage engine is described that acts as a source of truth for all data and handles versioning, access control etc. The dataset transformation mechanism is a key part to generate a dataset (snapshot) to serve different purposes. The described techniques can support different workflows, pipelines, or data orchestration needs, e.g., for training and/or evaluation of machine learning models.